\documentclass[12pt]{spieman}  
\usepackage[]{aas_macros}  
\usepackage{hyperref}
\usepackage{amsmath,amsfonts,amssymb}
\usepackage{graphicx}
\usepackage{setspace}
\usepackage{tocloft}
\usepackage{wrapfig}
\usepackage[symbol]{footmisc}

\def\b1{\boldsymbol{1}}

\def\ba{\boldsymbol{a}}

\def\bc{\boldsymbol{c}}

\def\bD{\boldsymbol{D}}

\def\bE{\boldsymbol{E}}

\def\rm{\mathrm{m}}

\def\bx{\boldsymbol{x}}
\def\by{\boldsymbol{y}}

\def\bz{\boldsymbol{z}}

\title{Instrumental Polarization in Stellar Coronagraphy: Coherent Behavior and its Implications for Dark Hole Optimization}

\author[$\dagger$]{Richard A. Frazin}
\affil[$\dagger$]{Dept. of Climate and Space Sciences, University of Michigan, Ann Arbor, MI 48109}

\cftpagenumbersoff{figure}
\cftpagenumbersoff{table} 
\begin{document} 
\maketitle


\begin{abstract}

Stellar coronagraphs designed for high-contrast imaging of exoplanets inevitably introduce a small amount of instrumental polarization, called \emph{secondary polarization}. At the contrast levels required to detect and characterize terrestrial planets, these effects may become significant. Instrumentally induced polarization is often referred to as ``incoherent," yet this use of the term lacks rigor. This work uses Jones calculus and vector field simulations, including interactions with dielectric surfaces to show that the secondary polarization is fully coherent with the input field, but it does not interfere with it due to orthogonality.  A key consequence of the coherence secondary polarization is that the process of creating a dark hole in the primary polarization tends to also significantly mitigate the intensity corresponding to the secondary polarization, called the \emph{secondary intensity}, in the dark hole region.
This reduction of the secondary intensity may lead to relaxed polarization design requirements in future coronagraphs.  Additionally, if the contrast is sufficient to make the secondary intensity non-negligible, modulation schemes to separate the planet from the instrumental light need to account for the modulation of the secondary intensity.

\end{abstract}


{\noindent \footnotesize\textbf{$\dagger$}Richard Frazin,  \linkable{rfrazin\emph{\_at\_}umich.edu} }

\section{Introduction}\label{sec: intro}

Direct imaging of exoplanets is among NASA's top priorities in future missions, such as the Habitable Worlds Observatory (HWO).
The principal challenge of direct imaging is the fact that exoplanets are located in close angular proximity to their vastly brighter host stars. 
Quantitatively, this means that the telescopic optical system must be able to achieve a planet-to-star brightness ratio (i.e., \emph{constrast}), of less than $10^{-8}$, perhaps even $10^{-11}.$\cite{LUVOIR_model_JATIS22, Mennesson2024_HWOlab}
Current designs to meet this daunting requirement feature stellar coronagraphs, which suppress on-axis light while allowing slightly off-axis beams to pass through relatively unimpeded.\cite{Cavarroc_IdealCoronagraph06}
Even assuming perfect optical surfaces, diffraction alone would put the contrast levels orders of magnitude above the aforementioned requirements.
Real surfaces have aberrations at high and low spatial frequencies that further degrade the contrast.\cite{Krist_End2End_Roman_JATIS23}

Achieving high contrast requires active wavefront control strategies, notably the use of deformable mirrors (DMs).
In the context of space-based systems, the most prominent approach involving DMs is a family of techniques known collectively as \emph{electric field conjugation (EFC)}.
EFC procedures use one or two deformable mirrors (DMs) to modulate the intensity measured in the image plane through alternating sensing and control steps.
The end result of EFC procedures is a region of the image plane called a  \emph{dark hole}  in which the starlight is suppressed to a high contrast level, in which one may hope to detect a planet.\cite{GiveonKern_EFC11,Kasdin_EFC16b,  Desai2024_EFClab}
In test bed settings, such procedures yield dark holes with contrasts of roughly $10^{-9}$.\cite{Belikov_LabDemo_SPIE22, Seo_JPLhiContrastResult_JATIS19, }
Any exoplanet light will be incoherent with the starlight. This incoherence is critical to the detection, as the central region of the exoplanet’s image will not be significantly modulated by the DM, provided the DM modulation does not appreciably degrade the Strehl ratio.\cite{Bottom_CDI_MNRAS17, Gladysz10}

Adding challenge to the situation are the nefarious effects of instrumental polarization, which arise due to reflection and refraction and cannot be avoided in optical systems that bring beams to a focus.
These effects on the polarization state of the light are called \emph{polarization aberration}.\cite{McGuire90, Breckinridge15, ashcraft2024SCoOB, Ashcraft2025_GSMT2}
In this article, the fields and the corresponding intensities associated with the instrument-induced polarization will be referred to as \emph{secondary} components, as opposed to the \emph{primary} components that are treated when polarization effects are ignored.
Baudoz \emph{et al.} argued that polarization aberration had an effect of $\sim10^{-8}$ on the TDH2 bench.\cite{Baudoz_Goos-Hanchen_Imbert-Fedorov}
Further, polarization-dependent aberrations were part of a comprehensive study on the effect of aberrations in the Roman Space Telescope Coronagraph, indicating the polarization effects contribute about $3\times10^{-10}$ to the intensity (contrast units).\cite{Krist_End2End_Roman_JATIS23}

The secondary intensity is commonly referred to as ``incoherent."\cite{breckinridge2003polarization, guyon2009WFC_PIAA, Kasdin_EFC16, ashcraft2024SCoOB}
This parlance is incorrect, as this article makes clear.
While it is certainly true that the primary and secondary components do not interfere, subjecting the secondary components to the adjective "incoherent'' implies that its comportment is indistinguishable from that of the truly incoherent planetary emission. The reality is that the secondary fields  are fully coherent with the primary fields, which will be demonstrated via simple yet rigorous arguments in the next section.
A critical consequence of this coherence is that the secondary intensity is modulated by the DM. However, its modulation follows a different functional dependence on the DM commands than the primary intensity, necessitating separate accounting should it not be negligible.

Perhaps lending currency to the widespread, though erroneous, the use of the term ``incoherent" is that the DM command corresponding to the dark hole in the primary intensity, denoted by the vector $\bc_0$, emerges as the culmination of a delicate minimization. The secondary intensity at $\bc_0$, by contrast, lies farther from any such extremum.
As a result, small perturbations about $\bc_0$ influence the primary intensity more readily than they do the secondary intensity. 

\section{Coherence Properties of the Primary and Secondary Fields}\label{sec: Coherence}

This section begins by defining the term ``coherent" in the context of stochastic processes, and then applies this definition to analyze the primary and secondary intensities.

\subsection{Stochastic Definitions of ``Coherent" and ``Incoherent"}

The theoretical treatment of polarization and interference rests on the formalism of stochastic processes, of which we avail ourselves only the simplest elements.
The propagation of electric fields through many optical systems, including stellar coronagraphs, can be carried out one frequency (or wavelength) at a time, but accounting for polarization and interference phenomena requires a bit more care, which leads to the notion of \emph{quasi-monochromatic} sources.\cite{StatisticalOptics, Collett93}
A quasi-monochromatic electric field in the $\hat{\bx}$-direction (along the $x$ axis), centered on the frequency $\nu$ may be represented as:
\begin{equation}
\bE(t) = \hat{\bx} E_x a(t) \exp[j 2 \pi \nu t] \, ,
\label{eq: def quasi-monochromatic}
\end{equation}
where $E_x$ is a complex-valued constant, representing the amplitude and phase, and  
$a(t)$ is a complex-valued stochastic process, called an \emph{envelope function} that provides rapid modulation on a time-scale that is orders-of-magnitude smaller than any conceivable detector integration time.
Eq.~\eqref{eq: def quasi-monochromatic} is a good representation of an electric field that results from passing light arising from thermal source, such as a star, through a narrow-band filter.  
Two distinct thermal sources, even two points on the same tungsten filament in a light bulb, will have statistically independent envelope functions.
Indeed, let us consider two distinct points on luminous tungsten filament sitting behind a narrow-band filter, and let us take their envelope functions to be $a(t)$ and $b(t)$.
The functions $a(t)$ and $b(t)$ are subject to the following conditions:
\begin{align}
& \mathcal{P}[a(t),b(t' + \tau)]  = \mathcal{P}[a(t)] \mathcal{P}[b(t' + \tau)] \; , \forall \tau \label{eq: stat indep} \\
& \overline{a(t)}  = \overline{b(t)}  = 0. \label{eq: zero mean envelopes} \\
& \overline{a(t)a^*(t)} = \overline{b(t)b^*(t)} = 1 \label{eq: unity variance} \\
& \overline{a(t)b(t + \tau)}  = \overline{a(t)b^*(t + \tau)}  = \overline{a^*(t)b(t + \tau)} = 0 \; , \forall \tau  \label{eq: incoherent envelopes}
\, ,
\end{align}
where the $\mathcal{P}$ represents probability, the superscript $^*$ indicates complex conjugation and the overbar indicates a time-average operator, which for the purposes of this article is equivalent to taking the mean of a stochastic process.   
In reality, integration over a finite amount of time is needed for Eqs.~\eqref{eq: zero mean envelopes} through~\eqref{eq: incoherent envelopes} to be effectively realized.
We assume that the required integration times are much less than any currently possible detector frame rate (say, $10^{-4}\,$s).
Eq.~\eqref{eq: stat indep} states that the processes $a(t)$ and $b(t)$ are statistically independent.
Eq.~\eqref{eq: zero mean envelopes} states that envelope functions are zero-mean, and 
Eq.~\eqref{eq: unity variance} states that the envelope functions have a variance of unity.
Eq.~\eqref{eq: incoherent envelopes} defines \emph{incoherence} as the vanishing of all first-order cross-correlations between the two envelope functions. 
Eq.~\eqref{eq: incoherent envelopes} in fact follows from Eqs.~\eqref{eq: stat indep} and Eq.~\eqref{eq: zero mean envelopes}.

Equation~\eqref{eq: incoherent envelopes} is the definition of ``incoherence,'' and no other will suffice.
Eq.~\eqref{eq: unity variance} indicates that the function $a(t)$ $b(t)$ are both fully self-coherent.
Two envelope functions $a(t)$ and $c(t)$ with unity variance are \emph{fully coherent} if $\big|\overline{a(t)c^*(t + \tau)}\big| = 1 $ for some value of $\tau$, and \emph{partially coherent} if 
$0 < \mathrm{max}_\tau \big|\overline{a(t)c^*(t + \tau)} \big| < 1 \,$.

\subsection{Stochastic Fields and Their Intensities}

Since starlight is generally weakly polarized, let us consider the light to be unpolarized for simplicity.
The generalization to a beam with known Stokes parameters is mentioned below.
To model an polarized beam, we now take $a(t)$ and $b(t)$ to be the envelope functions associated with the $x$ and $y$ components, respectively, of the field from the \emph{same} thermal source.
This differs from the previous usage introduced before Eq.~\eqref{eq: stat indep}, where they referred to distinct sources.
Under this interpretation, Eqs.~\eqref{eq: stat indep} through~\eqref{eq: incoherent envelopes} hold as well.\cite{StatisticalOptics,Collett93}
An unpolarized beam has fields in the $\hat{\bx}$ and $\hat{\by}$ directions that are statistically independent, which can be expressed in a form similar to Eq.~\eqref{eq: def quasi-monochromatic}
\begin{equation}
\bE(t) = \big[\hat{\bx} E_x a(t) + \hat{\by} E_y b(t) \big]\exp(j 2 \pi \nu t) 
\, ,
\label{eq: unpolarized quasi-monochromatic}
\end{equation}
where $E_x$ and $E_y$ are complex-valued constants (with $|E_x|=|E_y|$ since the beam is unpolarized), and functions $a(t)$ and $b(t)$ satisfy Eq.~\eqref{eq: incoherent envelopes}.
One can find an analog to Eq.~\eqref{eq: unpolarized quasi-monochromatic} for a partially polarized beam with known Stokes parameters by using their definitions and solving for $|E_x|$, $|E_y|$, the mean phase difference between $E_x$ and $E_y$, and $|\overline{a(t)b^*(t)}|$, the latter of which must be nonzero if there is any circular polarization.

If an unpolarized beam were to encounter an optical system that introduces no differential delays comparable to or greater than the coherence time $\tau_\mathrm{c}$, defined as the smallest value of $\tau_\mathrm{c}$ such that  $\overline{|a(t)a^*(t + \tau)|} \lesssim 1/2 $, then field in the output plane is given by

\begin{equation}
\begin{pmatrix}
E_x'(t) \\
E_y'(t)
\end{pmatrix}
=
\begin{pmatrix}
J_{xx} & J_{xy} \\
J_{yx} & J_{yy}
\end{pmatrix}
\begin{pmatrix}
	E_x a(t) \\
	E_y b(t)
\end{pmatrix}  \, ,
\label{eq: Jones stochastic}
\end{equation}
where the harmonic term $\exp(j 2 \pi \nu t)$ has been amputated.  
The $2\times 2$ matrix containing complex valued quantities $ J_{xx}, \, J_{xy} , \,
J_{yx},\,\mathrm{and} \; J_{yy}$ is the celebrated Jones Matrix.\cite{IntroFourierOptics}
Carrying out the matrix-vector multiplication in Eq.~\eqref{eq: Jones stochastic} results in the following vector field in the output plane:
\begin{equation}
\bE'(t) = \underbrace{\hat{\bx}J_{xx} E_x a(t) + \hat{\by}J_{yy} E_y b(t)   }_{\text{primary fields}}  +
  \underbrace{\hat{\bx} J_{xy}E_y b(t)   + \hat{\by} J_{yx} E_x a(t)  }_{\text{secondary fields}} \, ,
\label{eq: Jones stochastic output}
\end{equation}
where the primary and secondary fields have been identified.
In Eq.~\eqref{eq: Jones stochastic output}, it is self-evident that the primary field component $\hat{\bx}J_{xx} E_x a(t)$ and the secondary field component $\hat{\by} J_{yx} E_x a(t) $ are, in fact, \emph{fully coherent} with each other since their time dependence is governed by the same function, $a(t)$, in addition to both having the same factor $E_x$.   
Of course, the same sentiment applies to the primary field component $\hat{\by}J_{yy} E_y b(t)$ and the secondary field component $\hat{\bx} J_{xy} E_y b(t) $.

\begin{wrapfigure}{l}{0.45\textwidth}
	\hspace{-5mm}
	\includegraphics[width=0.52\textwidth]{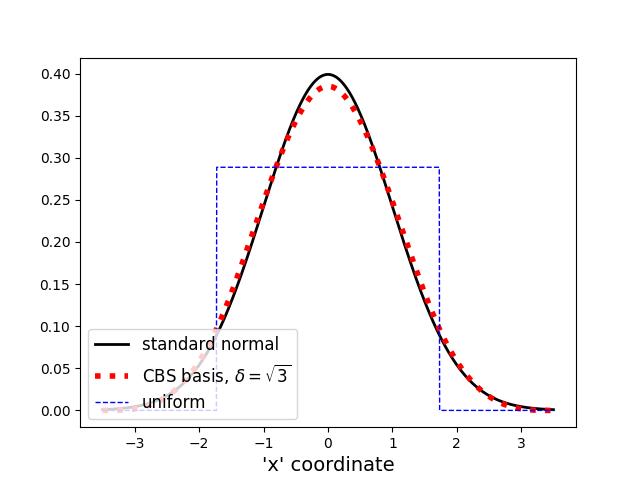}
	\caption{\small A cardinal B-spline (CBS) basis function in 1D with the width parameter $\delta$ set  to $\sqrt{3}$ and the univarate standard normal probability density function for comparison.  The uniform (pixel) basis functions is shown in addition.  All three functions integrate to unity and have a variance of unity as well.  Fig.~\ref{fig: Power Spectra} shows their power spectra.}\label{fig: CBS1D}
\vspace{-20mm}
\end{wrapfigure}
The intensity in the detector plane corresponding to Eq.~\eqref{eq: Jones stochastic output} is:
\begin{align}
I' & = \overline{\bE'(t) \cdot \bE'^*(t)}   \label{eq: detector intensity 0} \\
& = \underbrace{ |J_{xx} E_x |^2 + |J_{yy} E_y |^2  }_{\text{primary intensity}}  +
\underbrace{|J_{xy}E_y|^2   +  |J_{yx} E_x|^2  }_{\text{secondary intensity}} \, , \label{eq: detector intensity} 
\end{align}
where $\cdot$ represents the scalar (i.e., dot) product, and Eq.~\eqref{eq: unity variance} has been exploited.
The fact that the squaring and then taking the time average of the four terms in Eq.~\eqref{eq: Jones stochastic output} results in only four terms in Eq.~\eqref{eq: detector intensity} is due to the fact that $\hat{\bx} \cdot \hat{\by}=0$ and the incoherence conditions in Eq.~\eqref{eq: incoherent envelopes}.
At the risk of belaboring the point, the above discussion shows that coherence does not imply interference.

\section{Cubic Cardinal B-Spline Basis Function Expansions}\label{sec: CBS}

The spatial dimensionality of the fields in the previous section is intentionally ambiguous, but they can be thought of a corresponding to a single spatial point.
However, this study requires a spatial specification of the fields reflecting from the DM going into the coronagraph model.
The \emph{cubic cardinal B-spline} (CBS) basis function, plays a central role in the spatial model of the DM surface height as well as the phasor reflecting from the DM.

The CBS basis functions are not orthogonal, but they are commonly used in image processing and optics for interpolation of continuous functions.\cite{UnserSpline}
The CBSs are used for two purposes in this article:
Firstly, The height of the DM's membrane is modeled as an interpolation based on 2D CBS basis functions.  Secondly, the phasor corresponding to the continuous phase sheet created by the DM is approximated by another CBS expansion in order to represent the pupil plane field that is propagated through the optical system, as is explained in more detail later.  

The CBS basis function, has a width parameter $\delta$, and is given by the formula:\cite{UnserSpline}
\begin{equation}
\eta(x; \delta) = 
\begin{cases}
0, & \text{if } \left|\dfrac{x}{\delta}\right| \geq 2 \\
\dfrac{(2 - |x/\delta|)^3}{6}, & \text{if } 1 \leq \left|\dfrac{x}{\delta}\right| < 2 \\
\dfrac{2}{3} - \left(\dfrac{x}{\delta}\right)^2 + \dfrac{1}{2}\left(\dfrac{x}{\delta}\right)^3, & \text{if } 0 \leq \left|\dfrac{x}{\delta}\right| < 1 \, .
\end{cases}
\label{eq: CBS 1D}
\end{equation}
Thus, $\eta(x; \delta)$ has compact support (i.e., is nonzero) only in the interval $(-2 \delta, 2 \delta)$.
\cite{UnserSpline}
 The noteworthy properties of this function also include:
\begin{align}
\frac{1}{\delta}\int_{-2 \delta}^{2\delta} \eta(x;\delta) \mathrm{d} x & =  1 \, \: \: \: \mathrm{and} 
\label{eq: Unser unity}\\
\frac{1}{\delta}\int_{-2 \delta}^{2\delta} \eta(x;\delta)x^2 \mathrm{d} x & =  \frac{\delta^2}{3}  \,.
\label{eq: Unser variance}
\end{align}

\begin{wrapfigure}{l}{0.55\textwidth}
	\hspace{-8mm}
	\includegraphics[width=0.65\textwidth]{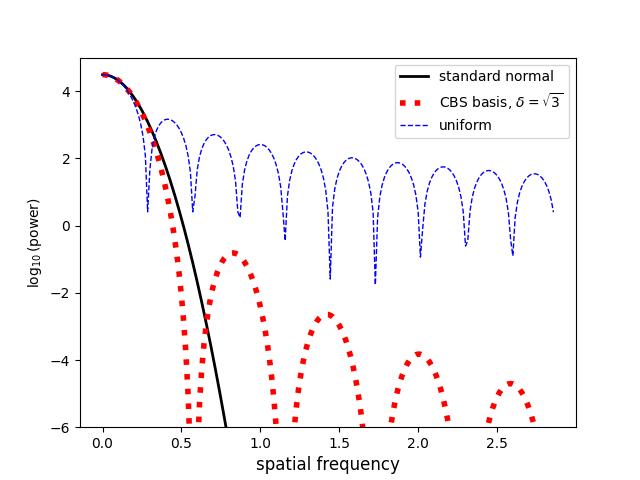}
	\caption{\small Power spectra of the functions shown in Fig.~\ref{fig: CBS1D}.  Note the sidelobes of the CBS function are orders-of-magnitude smaller than those of the uniform basis function.}\label{fig: Power Spectra}
\end{wrapfigure}

The function $(1/\sqrt{3})  \eta(x; \sqrt{3})$ bears a striking resemblance to the univariate standard normal probability density function (PDF), as depicted in Fig.~\ref{fig: CBS1D}, which also includes the corresponding uniform PDF, which can also serve as the pixel basis function.
The Fourier transform of the CBS basis function has sidelobes that are orders-of-magnitude smaller than the pixel basis function with the same area and variance, as depicted in Fig.~\ref{fig: Power Spectra}, which makes it advantageous for optical propagation.   

The two dimensional (2D) version of this CBS basis function is simply the tensor product of the one-dimensional function with itself; so, the 2D version is:
\begin{equation}
\eta(x,y;\delta) = \eta(x;\delta) \eta(y;\delta) \, .
\label{eq: CBS 2D}
\end{equation}

\begin{figure}[ht]
	\hspace{-17mm}
	\includegraphics[width=1.15\textwidth]{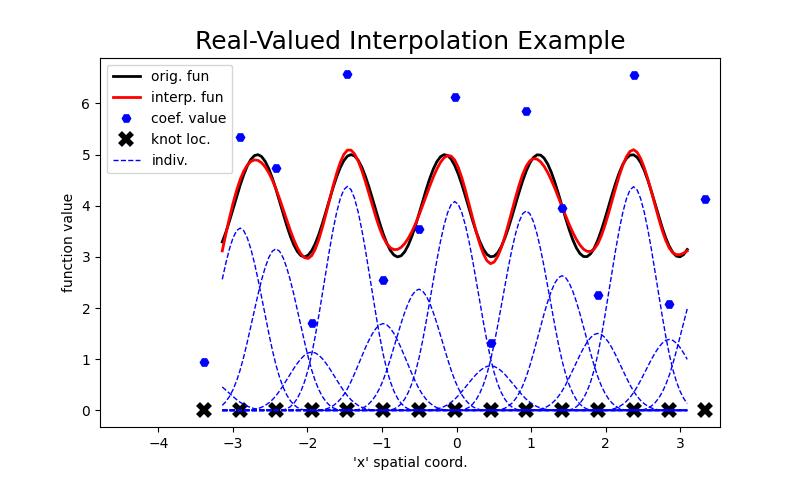}
	\caption{\small A CBS expansion, as per Eq.~\eqref{eq: CBS expansion 1D}, of the function
$f(x) = 4 + \cos(5x + \pi/4 )\, , \; -\pi < x < \pi$, with a CBS interpolation consisting of 15 basis functions.  The solid black curve is $f(x)$, the solid red curve is the interpolated value, and the dashed blue curves are the basis functions multiplied by their respective coefficients.  The blue dots are the coefficient values and the black ``x'' marks are the knots.}
	\label{fig: Interp1D}
\end{figure}

For the sake of clarity and brevity, the analysis below will be carried out in one spatial dimension (1D), since the extension from 1D to 2D is generally straightforward.
Following arguments similar to the usual sampling theory,\cite{UnserSpline} a band-limited function $f(x)$ can be approximated by a sum of CBS functions:
\begin{equation}
f(x) \approx \sum_k a_k \, \eta(x - k \delta; \delta) \, ,
\label{eq: CBS expansion 1D}
\end{equation}
where the values $\{ a_k \}$ are coefficients, and locations $\{ k \delta \}$ are known as the \emph{spline knots} or simply \emph{knots}.
An example of such an interpolation is depicted in Fig.~\ref{fig: Interp1D}.
Since the functions $\{  \eta(x - k \delta; \delta)  \}$ are not orthogonal, the coefficients are typically determined via least-squares regression.  However, for simulation of the DM surface height in this article, the coefficients are taken to be the DM commands themselves (see below).  

\begin{figure}[ht]
	\vspace{-30mm}
	\hspace{-10mm}
	\includegraphics[width=1.1\textwidth]{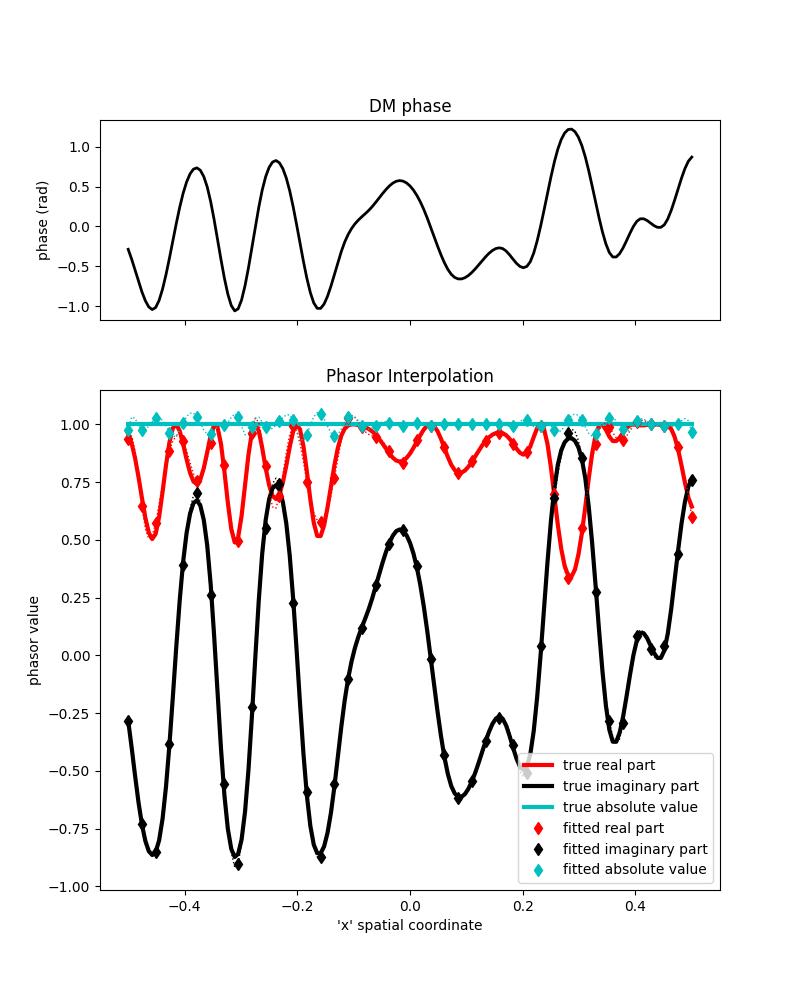}
	\caption{\small \emph{Top:} A DM height plot in phase units, with the 21 actuators given normally distributed random phases with a standard deviation of 1 radian.
	\emph{Bottom:} A comparison of the true phasor values to those of the fitted CBS interpolator.  The interpolator has 33 basis functions.  The solid cyan curve has a value of unity and corresponds to the absolute value of the true phasor.  The cyan diamonds correspond to the absolute value of the interpolator.  The solid red curve corresponds to the real part of the true phasor, and the red diamonds correspond to the real part of the interpolator.  The solid black curve is the imaginary part of the true phasor, and the black diamonds correspond to the imaginary part of the interpolator.  The interpolator closely matches the true phasor in this example.}
	\label{fig: Phasor Interp LowV}
\end{figure}
\begin{figure}
	\vspace{-25mm}
	\hspace{-10mm}
	\includegraphics[width=1.1\textwidth]{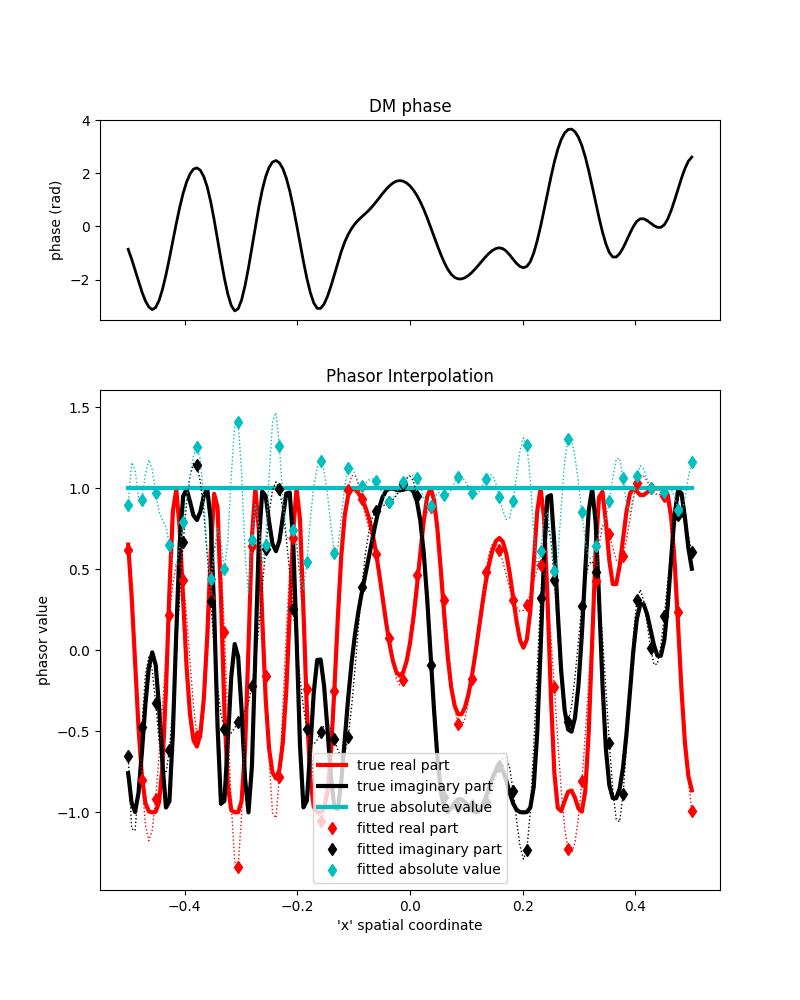}
	\caption{\small Identical to Fig.~\ref{fig: Phasor Interp LowV}, except DM height in the figure above has been multiplied by $\pi$.  Unlike the example in Fig.~\ref{fig: Phasor Interp LowV}, 33 basis functions are not sufficient to accurately approximate this phasor due its larger phase gradient.}
	\label{fig: Phasor Interp HighV}
\end{figure}
\clearpage

Numerical experiments indicate that, when approximating a function over the interval \linebreak \( (-L/2,\, L/2) \) using \( K \) terms in the sum of Eq.~\eqref{eq: CBS expansion 1D}, it is effective to choose an odd value for \( K \), set \( \delta = L / (K - 2) \), and place the first knot at $x=
- \delta \left( \frac{K - 3}{2} \right) \, $.
This configuration positions the first and last knots slightly outside the interval boundaries, thereby preventing the reconstructed function from artificially vanishing at the endpoints.
According to the Nyquist criterion, the configuration described above allows for the accurate approximation of functions containing power at spatial frequencies up to about $(K - 2)/(2L) $.

\subsection{Representation of the DM Phasor}\label{sec: DM}

Let $\lambda$ be the wavelength of interest and let $h'(x)$ be the height of the DM's surface at the point $x$.  Define $h(x) = 4 \pi h'(x)/\lambda$, which is the corresponding phase assuming normal incidence.
Henceforth, the height of the DM will be understood to be the phase it imparts to the impinging wavefront, i.e. the height is $h(x)$ not $h'(x)$.
Typically [e.g., \citenum{Kasdin_EFC16b}], the height of the DM is modeled as a linear combination of Gaussian \emph{influence functions}- a nomenclature that suggests each actuator of the DM has a spatially extended region of influence.  
The DM surface model in this article is scarcely different; instead of taking the influence functions to be Gaussians, the CBS basis functions are used instead.  Given the similarity of the of CBS basis function and the Gaussian shown in Fig.~\ref{fig: CBS1D}, one would expect the CBS basis functions to perform comparably in this role.  

Following Eq.~\eqref{eq: CBS expansion 1D}, the height (in phase units) of the DM is given by the sum:
\begin{equation}
h(x) = \sum_{k=0}^{K-1} c_n \eta(x - k \delta_p; \, \delta_p)  \, ,
\label{eq: DM expansion 1D}
\end{equation}
where $\delta_p$ is the distance between actuators, known as the \emph{pitch}, and $\bc = (c_0,\, \dots, \, c_{K-1})$ is vector of DM commands and K is the number of actuators.
Since $h(x)$ is already a phase value in radian units, the corresponding phasor function is 
\begin{equation}
u'(x) = \exp \left[ j h(x)   \right] = \exp \left[ j \sum_{k=0}^{K-1} c_n \eta(x - k \delta_p; \, \delta_p) \right] \, .
\label{eq: DM phasor}
\end{equation}

To represent a plane wave wave interacting with the DM and then passing through the coronagraph, one could simply multiply the field corresponding to the plane wave by $u(x)$ in Eq.~\eqref{eq: DM phasor} and propagate it through the optical system.
However, the matrix-based coronagraph model, explained in more detail in Sec.~\ref{sec: Jones Rep}, used here requires the input field to be expressed in terms of CBS coefficients.
Thus, we consider a CBS expansion of $u'(x)$:
\begin{equation}
u(x) = \sum_{n=0}^{N-1} a_n \eta(x - n\delta; \, \delta)   \approx u'(x) \, ,
\label{eq: CBS phasor}
\end{equation}
where $N$\ is the number of complex-valued coefficients, placed into the vector $\ba = (a_0, \, \dots, \, a_{N-1}) $, expected by the coronagraph model and $\delta$ is distance between knots in the coronagraph model.  
The coefficients $\ba$ in Eq.~\eqref{eq: CBS phasor} are determined via a least-squares fit to $u'(x)$, whereas the DM command $\bc$ in Eq.~\eqref{eq: DM expansion 1D} is assumed to be given.

While Eqs.~\eqref{eq: DM expansion 1D} through~\eqref{eq: CBS phasor} are not fettered by linearization in the DM command, ensuring the accuracy of the expansion in Eq.~\eqref{eq: CBS phasor} requires some attention. 
In particular, the amplitude of the DM phasor $u'(x)$ is unity, but $u(x)$ may depart from this ideal.
Numerical experiments show that the larger the gradients in the height $h(x)$, the larger $N$ must be to ensure an accurate approximation (note: $\delta \propto 1/N$).
In this article, $N$ is fixed by the coronagraph model, and this translates to a constraint on the DM command.  
Figs.~\ref{fig: Phasor Interp LowV} and \ref{fig: Phasor Interp HighV} illustrate this concern.
The top panel in both figures is the phase imparted by the DM, modeled as a sum of 21 CBS basis functions.
The phase in Fig.~\ref{fig: Phasor Interp HighV} is $\pi$ times that in Fig.~\ref{fig: Phasor Interp LowV}.  
The lower panel in these figures is a comparison between between the true phasor, i.e., the exponentiated phase in the upper panel, and its representation in terms of 33 CBS basis functions with complex-valued amplitudes, as indicated in Eq.~\eqref{eq: CBS phasor}.
It is rather evident the CBS expansion with 33 basis functions is far more successful with the smaller phase gradients in Fig.~\ref{fig: Phasor Interp LowV} than it is with the larger phase gradients in Fig.~\ref{fig: Phasor Interp HighV}.

\begin{figure}[t]
	\centering
	\includegraphics[width=1.1\textwidth]{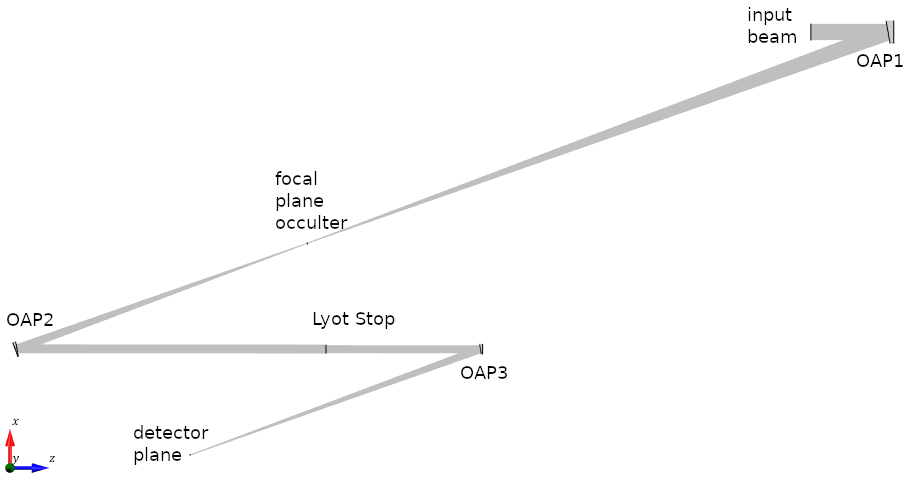}
	\caption{\small A schematic diagram of the Lyot-type stellar coronagraph used in these simulations. The DM, which is not shown, modulates the otherwise collimated input beam before OAP1. The $y$ axis is invariant with this design.}
	\label{fig: coronagraph}
\end{figure}

\section{The Coronagraph Model}\label{sec: Layout}

The CBS representation of the DM phasor described in the previous section enables the matrix-based coronagraph model, whose details are provided in this section.
The function of this model is propagate the DM phasor to the detector plane.
The simulation results in this article correspond to a square Lyot-type stellar coronagraph shown schematically in Fig.~\ref{fig: coronagraph} and with key parameters summarized in Table~\ref{table: optical system}.
This coronagraph consists of 3 OAPs, an occulter in the initial focal plane, and a Lyot stop in the plane corresponding to Fourier transform of the field in the initial focal plane.
The initial propagation direction of the beam before it encounters OAP1 is in the $+z$ direction.
The simulation is configured so that the $y$ axis is invariant, so the principal axis of the beam remains in the $xz$ plane after each OAP encounter.

\begin{figure}[t]
	\hspace{-5mm}
	\begin{tabular}{r l}
		\includegraphics[width=0.5\textwidth]{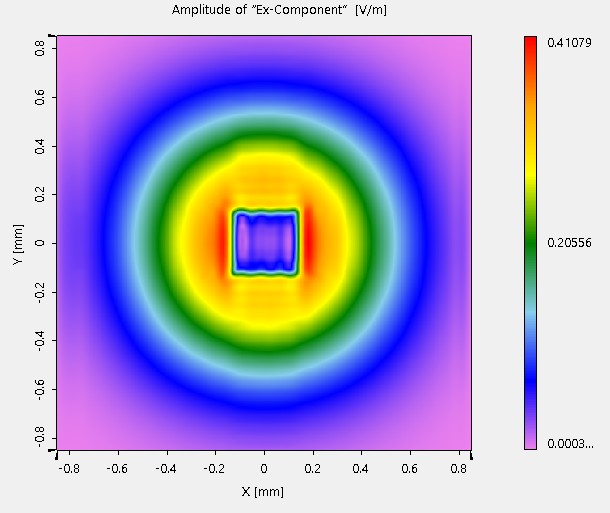}	&
		\includegraphics[width=0.5\textwidth]{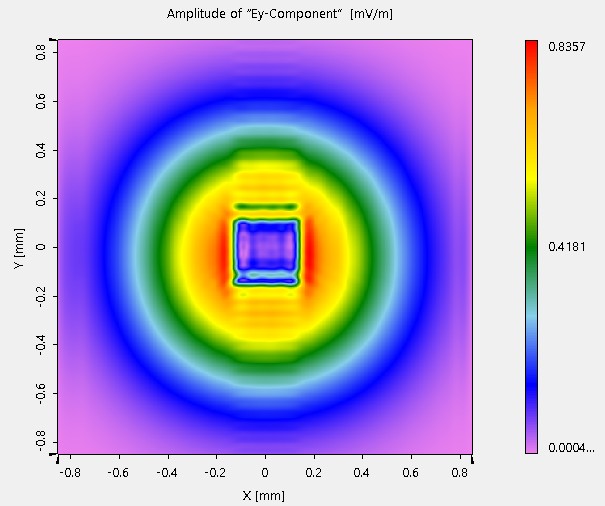} \\
		\includegraphics[width=0.5\textwidth]{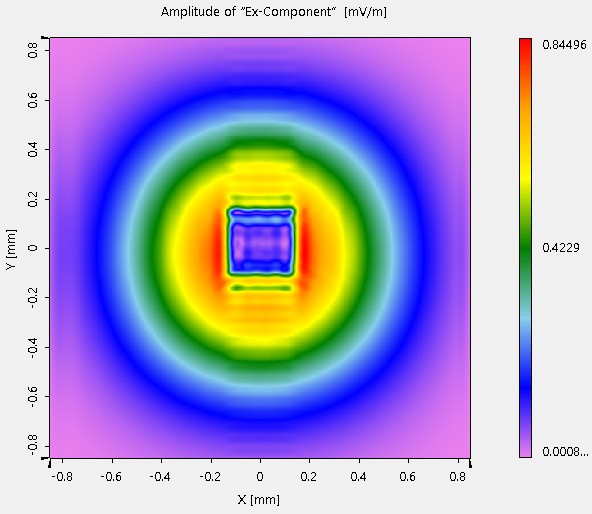}	&
		\includegraphics[width=0.5\textwidth]{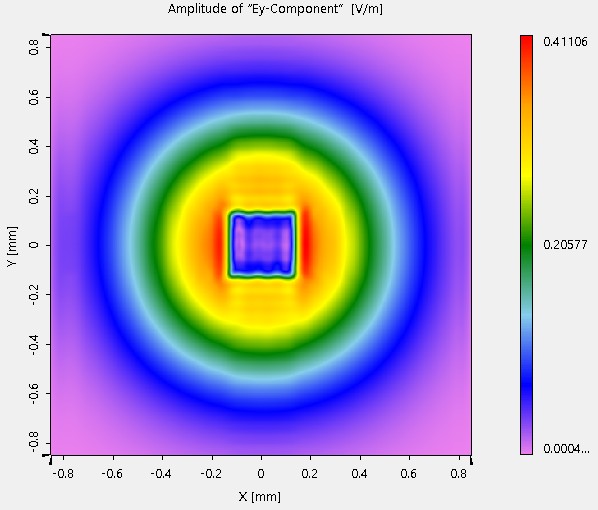} 
	\end{tabular}
	\caption{\small The absolute values of the columns of the $\bD_{xx}$ (\emph{upper left}), $\bD_{yy}$ (\emph{lower right}),  $\bD_{xy}$ (\emph{upper right}), and$\, \bD_{yx}$ (\emph{lower left}) matrices associated with the CBS basis function centered at $(3\delta,\, 14\delta)$, where $\delta$ is the grid spacing of the spline knots in the entrance pupil.  The columns have been reshaped into images corresponding to the detector plane.  Since the spline knots locations range between $\pm 16 \delta$ in both the $x-$ and $y-$ directions, this particular knot is located slightly to the right of the center but near the top of the entrance pupil.  The quantities displayed in the upper left and lower right are identical (apart from a negligible scale factor---see text), while those in the upper right and lower left have small differences that can be seen with careful visual inspection.}\label{fig: Jones spline}
\end{figure}

\subsection{Jones Representation}\label{sec: Jones Rep}

This section presents the Jones model of the coronagraph- the spirit of which has been portended in Eq.~\eqref{eq: Jones stochastic}.
While the quantities $ J_{xx}, \, J_{yy} , \, J_{xy},\,\mathrm{and} \; J_{yx}$ in Eq.~\eqref{eq: Jones stochastic} are complex-valued scalars, their analogues that describe the vector-field response of the coronagraph are complex-valued matrices.
Specifically, the matrices $\bD_{xx}$, $\bD_{yy}$,  $\bD_{xy}$, and$\, \bD_{yx}$ are $M\times N$, where $M$ is the number of detector pixels and $N$ is the number of spline coefficients used to specify the input field, in this case $N=33\times33=1089$.

The matrices $\bD_{xx}$ and $\bD_{xy}$ represent the response to $x-$polarized input, and $\bD_{yy}$ and $\bD_{yx}$ represent the response to $y-$polarized input.
Assuming the input is purely $y-$polarized, column $k$ of $\bD_{yx}$ is the $x-$component of the field in the detector when the field in the entrance pupil is zero, apart from the $k$\underline{th} (using 1D indexing) CBS basis function having unit amplitude and zero phase.
Perhaps not unexpectedly, the matrices $\bD_{xx}$,  $\bD_{xy}$, and$\, \bD_{yy}$ are defined in an analogous manner.
The matrices $\bD_{xx}$, $\bD_{yy}$,  $\bD_{xy}$, and$\, \bD_{yx}$ are constructed by propagating each CBS basis function through the coronagraph.
Fig.~\ref{fig: Jones spline} is an example of this for the CBS basis function centered at $(3\delta,\, 14\delta)$ (so, the 1D index is $k=976$, for a $33\times33$ grid), where $\delta$ is the grid spacing of the spline knots in the entrance pupil.

Following Sec.~\ref{sec: DM}, a DM command vector $\bc$ results in vector of spline coefficients $\ba$.
Then the vector field in the detector plane is given by a generalization of Eq.~\eqref{eq: Jones stochastic output} to the coronagraph model. For an unpolarized input beam, this is:
\begin{equation}
\bE(t) = \underbrace{\hat{\bx}\bD_{xx} \ba a(t) + \hat{\by}\bD_{yy} \ba b(t)   }_{\text{primary fields}}  +
\underbrace{\hat{\bx} \bD_{xy} \ba b(t)   + \hat{\by} \bD_{yx} \ba a(t)  }_{\text{secondary fields}} \, ,
\label{eq: Jones stochastic output big}
\end{equation}
where the vector $\bE'(t)$ is defined somewhat usually to have a length $2M$ in order to contain both the $x$ and $y$ components of the electric field at each of the $M$ detector pixels.
Following logic that is similar to that used to go from Eq.~\eqref{eq: Jones stochastic output} to Eq.~\eqref{eq: detector intensity}, we find that the intensity corresponding to Eq.~\eqref{eq: Jones stochastic output big} is:
\begin{align}
I' & = \overline{\bE'(t) \cdot \bE'^*(t)}   \label{eq: detector intensity 0} \\
& = \underbrace{ |\bD_{xx} \ba |^2 + |\bD_{yy} \ba |^2  }_{\text{primary intensity}}  +
\underbrace{|\bD_{xy} \ba|^2   +  |\bD_{yx} \ba|^2  }_{\text{secondary intensity}} \, , \label{eq: detector intensity big}
\end{align}
\begin{wraptable}[16]{l}{0.55\textwidth}
	\begin{tabular}{|c|c|}
		\hline
		Parameter & Value \\
		\hline
		wavelength ($\lambda$) & $1 \, \mu$m \\
		input beam diameter ($D$) & 19.34 mm \\
		effective  $f$\# & 44 \\
		DM actuators (spanning $D$) & $21 \times 21$ \\
		occulter opaque diameter & $240 \, \mu$m \\
		occulter transiton edge width & $40 \, \mu$m \\
		Lyot stop clear diameter & $9.1 \,$mm \\
		Lyot stop transition edge width & $1.2 \,$mm \\
		OAP1 focal length ($f$) & 800 mm \\
		OAP2, OAP3 focal length ($f$) & 400 mm \\
		OAP1, OAP2, OAP3 off-axis angle ($\zeta$) & $20^\circ$ \\
		SiO$_2$ thickness on Ag OAP mirrors & $470\,$nm \\
		\hline
	\end{tabular}
	\caption{\small Lyot coronagraph parameters used for the simulations.}
	\label{table: optical system}
\end{wraptable}
where it is assumed that input is unpolarized.
Recalling that the vector of CBS coefficients $\ba$ is found from the DM command (see Sec.~\ref{sec: DM}), we see that the primary and secondary intensities are the fully coherent results of the same DM command, and might be expected to manifest similar characters.

The physical layout of the coronagraph, provided above in Sec.~\ref{sec: Layout}, results in a $y$ axis that is invariant throughout the optical system.
This makes the notation convenient because if the input field is purely polarized in the $x-$direction, the primary field in the detector plane will also be in the $x-$direction, which requires the secondary field to be in the $y-$direction.
Thus, the matrices $\bD_{xx}$ and $\bD_{yy}$ provide the primary responses of the coronagraph, while $\bD_{yx}$ and $\bD_{xy}$ provide the secondary responses.
The color scales in Fig.~\ref{fig: Jones spline} indicate that, at least for the basis function located at $(3\delta,\, 14\delta)$, the magnitudes of the primary responses are factor of about 500 greater than those of the secondary responses, which corresponds to a weakly polarizing system.

\begin{figure}[t]
	\hspace{-12mm}
	\begin{tabular}{r l}
		\includegraphics[width=.55\textwidth]{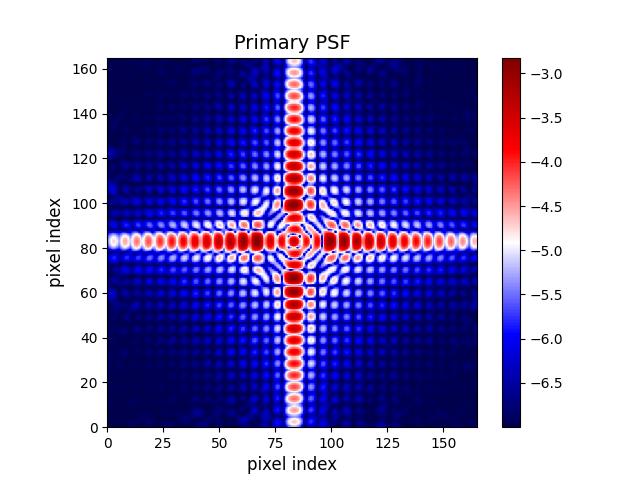} & 
		\includegraphics[width=.55\textwidth]{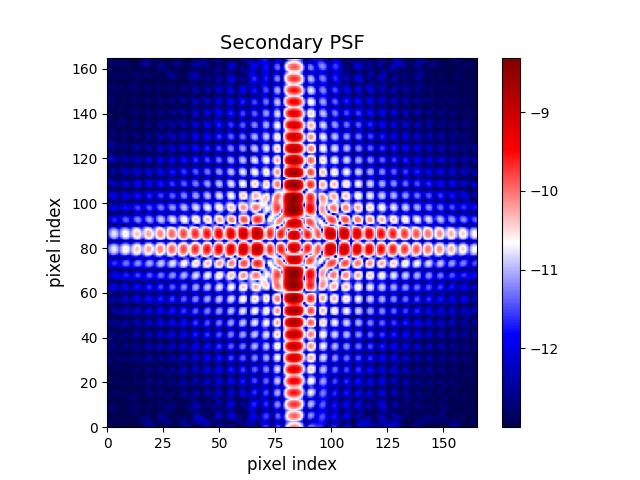}
	\end{tabular}
	\caption{\small PSFs from coronagraph simulations.
		\emph{Left:}  Primary PSF (contrast units). The color scale obtains its minimum value at $10^{-7}$ for display purposes.  
		\emph{Right:}  Secondary PSF (contrast units).  The color scale obtains its minimum value at $10^{-13}$ for display purposes.
	}
	\label{fig: PSFs}
\end{figure}

\subsection{Fourier Optics, Symmetries and Non-paraxiality}\label{sec: symmetry}

Fourier optics modeling typically treats the electric field as a scalar quantity and does not include polarization effects.
The scalar field $u$ in Fourier optics modeling corresponds to $u = (E_x + E_y)  /\sqrt{2}$ (assuming unpolarized light) and only treats the dominant response, which is assumed to be same for both the $x$ and $y$ components.  
This is consistent with Jones coronagraph model in this article.  
Indeed, in Fig.~\ref{fig: Jones spline}, the dominant responses (upper left and lower right) are imperceptible, but the color scales differ by about 1 part in 1500.
This slight difference is caused by a slight difference in the reflection efficiency in the OAP mirrors for the two polarizations (see Fig.~\ref{fig: OAP coating}).
Indeed, for this optical system $\bD_{xx} = \bD_{yy}$ (apart from a negligible scale factor) and could be calculated within a scalar Fourier optics framework.

Besides the relation $\bD_{xx} = \bD_{yy}$, the coronagraph simulated here exhibits another approximate symmetry: $| \bD_{xy} | \approx |\bD_{yx}|$, in part due to invariance of the $y$ axis in the optical system layout described above.
This trait is on display in Fig.~\ref{fig: Jones spline} where the magnitudes of the secondary responses (lower left and upper right) are difficult, but not impossible, to distinguish visually and the color scales differ by roughly $1\%$.
Applying the relations $\bD_{xx} = \bD_{yy}$ and $| \bD_{xy} | \approx |\bD_{yx}|$ to Eq.~\eqref{eq: detector intensity big} and dropping a factor of 2 for aesthetic purposes, the detector intensity is given by
\begin{equation}
I' \approx \underbrace{ |\bD_{xx} \ba |^2   }_{\text{primary intensity}}  +
\underbrace{|\bD_{xy} \ba|^2  }_{\text{secondary intensity}} \, .
\label{eq: detector intensity symmetric} 
\end{equation}

Setting the values of the coefficient vector $\ba$ to unity corresponds to a flat DM, and the primary and secondary intensities obtained are the PSFs shown in Fig.~\ref{fig: PSFs}---we call these the \emph{nominal} primary and secondary PSFs.
Were it not for the symmetries mentioned just above, there would be two primary PSFs and two secondary PSFs.  
The color scale in Fig.~\ref{fig: PSFs} corresponds to contrast units, which are set by scaling the maximum intensity of a slightly off-axis source (at an angle of $5 \, \lambda/D$ to skirt the occulter in the first focal plane) to unity. 

Simulations not provided here show that when a plane wave corresponding to a source that is far off-axis with an angle of, say, $\theta_\mathrm{offax} = 150 \lambda/D$ or more, is placed into the entrance pupil, the resulting dominant fields may differ rather significantly due to non-paraxial effects.
The physical optics methods described in Sec.~\ref{sec: prop methods} accurately capture non-paraxial effects that cannot be modeled using the common Fresnel or Fraunhofer Fourier optics approximations.\cite{IntroFourierOptics}
Note the matrices $\bD_{xx}$ and $\bD_{yy}$ (as well as the other two siblings) in these simulations are constructed with only $33\times33$ resolution of the input and therefore have a Nyquist frequency of $\theta_\mathrm{offax} \approx 15 \lambda/D$.  
Thus, their input spatial resolutions are insufficient to represent significantly paraxial beams.
To perform such calculations, the VLF coronagraph model is fed an inclined plane wave instead of a CBS basis function.

\section{Propagation Methods}\label{sec: prop methods}

While this coronagraph design is simple, the propagation techniques applied to the vector field are state-of-the-art.
As discxussed in Sec.~\ref{sec: Jones Rep}, calculating the matrices $\bD_{xx}, \, \bD_{xy}, \, \bD_{yy}\;$and $\bD_{yx}$ requires propagating the CBS basis functions through the coronagraph. 
This task is performed using the VirtualLab Fusion (VLF) optical simulation package developed by LightTrans, GmbH.
VLF is a comprehensive physical optics software platform widely used in photonics and microscopy [e.g., \citenum{Wyrowski_DiffMetaLens_SPIE19,  Wyrowski_NanoOpt_JOSAA2020}].
It supports a broad range of algorithms for field propagation and surface interaction, enabling self-consistent treatment of vectorial, geometrical, and diffractive phenomena.
VLF's algorithms do not rely on the assumption of paraxiality, which is implicit in the Fresnel and Fraunhofer approximations,\cite{IntroFourierOptics} unless explicitly chosen by the user.
VLF always represents the electromagnetic field at every stage of the model without ever resorting to a ray-based representation, as is standard with ray tracing engines.\footnote{VirtualLab Fusion (VLF) supports modeling of magnetic materials and thus retains the magnetic field vector by default. However, in scenarios involving only non-magnetic media-- as is the case in this article-- the electric field alone is sufficient, and the corresponding magnetic field can be obtained via the Maxwell-Faraday equation in $\boldsymbol{k}$-space.}

VLF's field-based representation allows algorithms that are either geometrical (denoted as \emph{geometrical algorithms}) or non-geometrical (denoted as \emph{non-geometrical algorithms}).
To put it crudely, one might say that in the geometrical regime, light at any given point travels in a single direction.
Non-geometrical algorithms incorporate diffractive effects incurred by the field as it propagates from one surface to the next, whereas the operation of geometrical algorithms on the field does not cause additional diffraction while preserving the diffractive structure already present.
Since both the geometrical and non-geometrical algorithms in VLF operate on the electromagnetic field, optical system models can take advantage of a seamless transition between geometrical and non-geometrical regimes, which is indeed required to calculate the polarization aberrations in the OAP-based coronagraph simulated in this article.\cite{Wyrowski_PolarizFieldTr_SPIE2019, Wyrowski_Seamless_SPIE24}

\begin{figure}[t]
	\hspace{-3mm}
	\begin{tabular}{l l}
		\includegraphics[height=7.3cm]{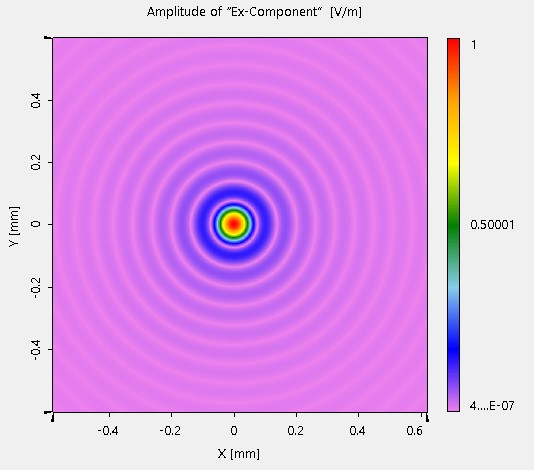} & \hspace{-6mm}
		\includegraphics[height=7.3cm]{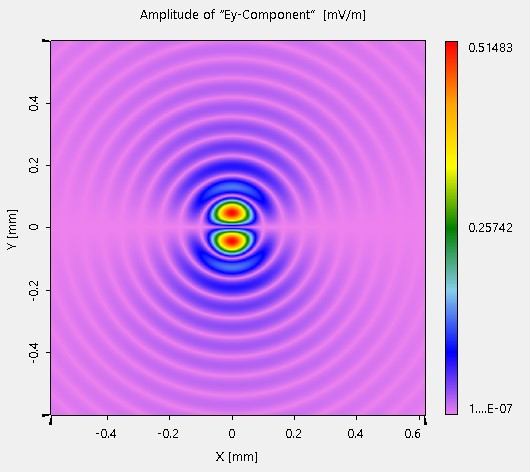} \\
		\includegraphics[height=7.3cm]{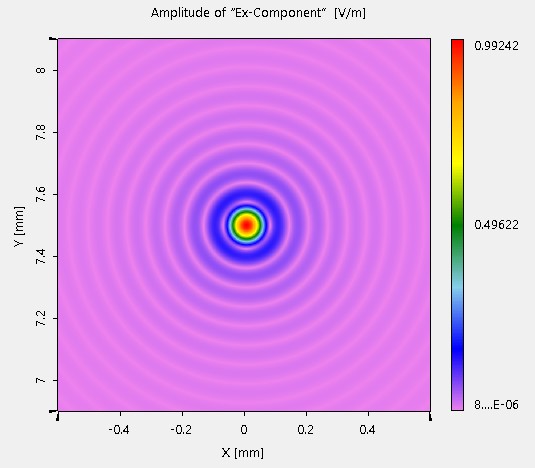} & \hspace{-6mm}
		\includegraphics[height=7.3cm]{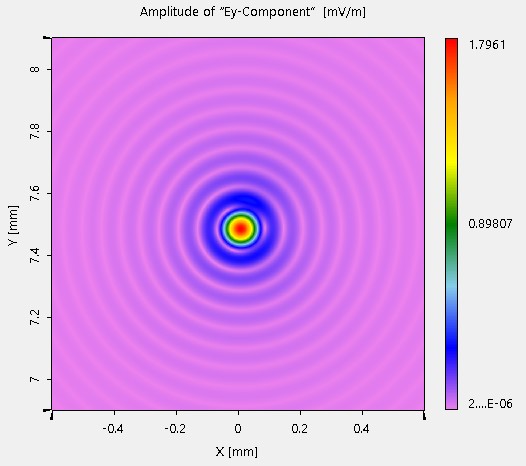} 
	\end{tabular}
	\caption{\label{fig: fields OAP}
		A non-coronagraphic simulation of focal plane $|$fields$|$ corresponding to a collimated beam focused by a single $f$\#~=~62.5 OAP with an off-axis reflection angle of $20^\circ$.  
		The units of the color scale are V/m in the left column and mV/m in the right column.
		The values are all on the same scale, but each panel has its own colorbar to make its features visible.
		The upper row corresponds to an on-axis input beam (focal point at the origin), while the lower row shows results for an off-axis input beam that makes an angle of $120\,\lambda/D\,$with the $y$ axis.
		The input beams in both cases are $x$-polarized.
		The two images on the left are primary polarization and the two on the right are secondary polarization.
		The upper-left image shows $|E_x|$ (peak~=$\, 1\,$V/m)  and the upper-right image shows $|E_y|$ (peak~=$ \, 0.5\,$mV/m).
		The lower-left image shows $|E_x|$ (peak~=$\, 0.992\,$V/m), and the lower-right image shows $|E_y|$ (peak~=$\,1.8\,$mV/m).
		Note the differences in scale and form of the secondary $|$fields$|$ due to the nonparaxial properties of the off-axis beam shown in the images on the right.
		There is no need to show results for $y$-polarized input beams, since they are symmetrical.
	} 
\end{figure} 

\clearpage

\subsection{Detailed OAP Treatment}

This section discusses the algorithms used by VLF to model the propagation through the OAPs. 
This is a comprehensive calculation including polarization aberration, which requires a non-paraxial and vectorial treatment due to the curved surface of the OAP.
While polarization ray-tracing (PRT) can be used to evaluate the change in the Stokes vector associated with a given ray reflecting from a given point on the OAP surface, making PRT consistent with diffraction phenomena remains an area of active research. 

Fig.~\ref{fig: fields OAP} shows $|E_x|$ and $|E_y|$ in the focal plane  for a collimated $x$-polarized beam 
\begin{wrapfigure}{l}{0.45\textwidth}
	\hspace{-11mm}
	\includegraphics[width=0.52\textwidth]{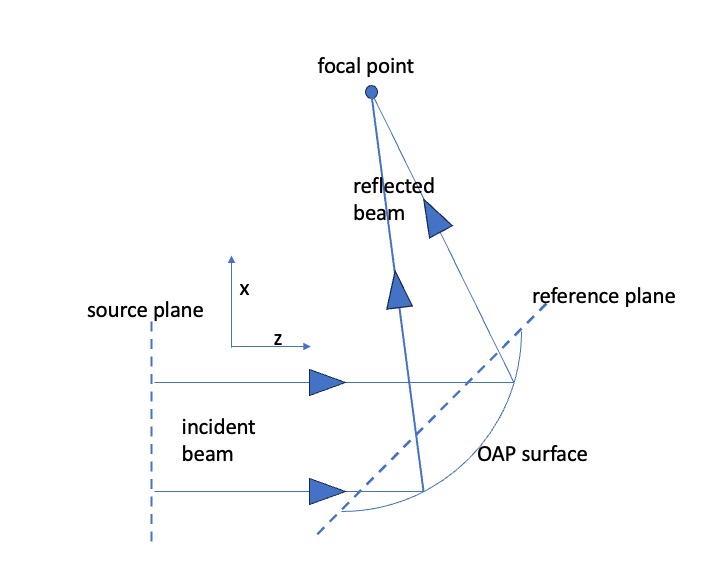}
	\caption{\small A sketch of an off-axis parabola (OAP) bringing a collimated beam to a focus for the purpose of explaining the roles of several of VLF's propagation algorithms. See~Table~\ref{table: prop methods}.  The angles and curvature of the surface are exaggerated for clarity.  }\label{fig: VLF_OAP}
	\vspace{-2mm}
\end{wrapfigure}
focused by an OAP mirror.  
The results for a $y-$polarized input beam are not shown because they are perfectly symmetrical, as discussed in Sec.~\ref{sec: symmetry}.    
The diameter of the collimated input beam is 8~mm, the wavelength of the light is $1 \, \mu$m, and the OAP's  off-axis reflection angle is 20$^\circ$ in the $xz-$plane, with a focal length of 500~mm, giving the system $f$\#~=~62.5.
The OAP material is silver with a thin glass coating, as per Table~\ref{table: optical system}.

The upper pair of images in Fig.~\ref{fig: fields OAP} corresponds to an on-axis, i.e., $+z$ direction, input beam.
The lower pair of images corresponds to an off-axis input beam.  For the off-axis beam, the initial propagation direction is $\hat{\by} \sin\eta + \hat{\bz}\cos \eta  $, where $\eta = - 120 \, \lambda/D = - 0.86^\circ$, which is why these images are centered at $(0, \, 7.5\, \mathrm{mm})$, rather than the origin.
The lower left image is similar to the upper left image-- minor differences can be seen upon close inspection.
The small difference in the peak values in the images on the left is mostly due to the angular dependence of the Fresnel reflection coefficients applied at the OAP surface.
The secondary $|\mathrm{field}|$ shown in the lower right image is rather different than in the upper right image, including a peak value of $|E_y|$ more than three times greater.  
Additionally, while the upper right image exhibits a bimodal behavior with symmetry about both the $x$ and $y$ axes, the lower image is not bimodal, nor is it symmetric about the $y$ axis.

To understand how VLF handles the OAPs, refer to Fig.~\ref{fig: VLF_OAP}, which is a sketch of a collimated beam specified in a source plane focused by a concave mirror. 
VLF propagates the vector field through an OAP via the following steps:
\begin{enumerate}
		\item{Propagate the incident vector field from the source plane to the \emph{reference plane}.  This is referred to as \emph{plane-to-plane} in Table~\ref{table: prop methods}.  The reference plane is defined to be the plane that is closest to the curved surface that is not obstructed.  The reference plane will likely be inclined relative to the source plane.}
	\item{Propagate the vector field from the reference plane to the curved surface of the mirror.  This is the field that is impinging on the dielectric surface, the calculation of which is referred to as \emph{plane-to-curved surface} in Table~\ref{table: prop methods}.}
	\item{At the mirror's surface, calculate the reflected vector field.
	This is referred to as \emph{reflection at a curved surface} in Table~\ref{table: prop methods}.}
	\item{Propagate the reflector vector field from the mirror's surface to the reference plane.
	This is referred to as \emph{curved surface-to-plane} in Table~\ref{table: prop methods}.}
	\item{Propagate the reflected vector field from the reference to the focal plane (or other output plane, if not being brought to a focus), taking the relative incliations of the two planes into account.
	This is referred to as \emph{plane-to-plane} in Table~\ref{table: prop methods}. }   
\end{enumerate}

\begin{table}[h]
	\hspace{-1mm}
	\begin{tabular}{| r| c | c | c | c |c|}\hline
		Propagation & Algorithm & Vector & Paraxial & Geometrical & Comment \\ 
		Task    & & field & approx.  & approx. & \\  \hline
		plane-to-plane & ASM w/ FFT & yes & no &no &  inclined planes OK\\ \hline
		plane-to-  & ASM w/ PFT  & yes &no & yes     & sub-Fresnel distance \\
		curved surface &   &   & & &  \\ \hline
		reflection at a& LPIA & yes & n/a & yes & includes TM/TE modes  \\ 
		curved surface & & & & & \\ \hline
		curved surface to & ASM w/ PFT &yes & no  & yes   & sub-Fresnel distance \\
		reference plane  &   & & &  & \\ \hline
	\end{tabular}
	\caption{Table of propagation tasks and the algorithms VLF employs in this article.  Refer to Fig.~\ref{fig: VLF_OAP} and see text for more details.  When ``geometrical approx.'' is true, the field incurs no additional diffraction in this propagation step.}\label{table: prop methods}
\end{table}

\begin{wrapfigure}[21]{l}{0.45\textwidth}
	\hspace{0mm}
	\includegraphics[width=0.45\textwidth]{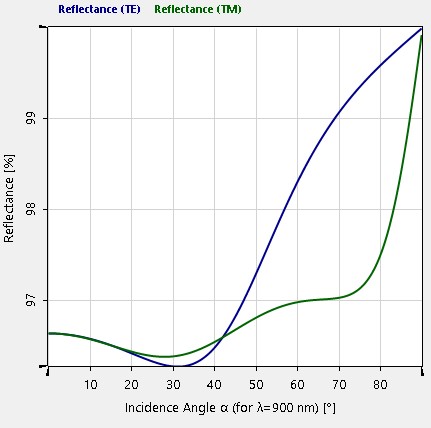}
	\caption{\small Magnitudes of the complex-valued TM and TE reflection coefficients at $\lambda=900\,$nm as a function of the angle of incidence for a silver film and a $470 \,$nm thick SiO$_2$ coating.}\label{fig: OAP coating}
	\vspace{-1mm}
\end{wrapfigure}
The propagation operations required for these simulations are summarized in Table~\ref{table: prop methods}, and the VLF implementation of these tasks is described below.

\noindent {\bf plane-to-plane:}  
One of principal tasks is free-space propagation of the vector field from one plane to the next.
In these simulations, free-space propagation between any two planes is carried out with the angular spectrum method (ASM),\cite{IntroFourierOptics} which relies on  the FFT algorithm to calculate the FT.
Crucially, VLF employs an ASM algorithm that allows propagation between non-parallel planes.\cite{Matsushima_NonParallelAngularSpectrum, zhang_NonParallelAngularSpectrum} 
The ASM is applied independently for each component of the vector field.

\noindent{\bf plane-to-curved surface:}
Referring to Fig.~\ref{fig: VLF_OAP}, one can see that after the incident field has been evaluated on the reference plane, it next needs to be propagated to the curved mirror surface.
The exaggerated curvature in Fig.~\ref{fig: VLF_OAP} may be misleading; in realistic coronagraph designs, the mirror curvature is gentle and the distance to the reference plane is small.
Indeed, in any coronagraphic imaging system, at least within the limitations of this author's imagination, the distance between the reference plane and the curved surface of an optic is tiny compared to any other relevant dimension.
Consider an OAP with a reflected focal length $f$ and a diameter $d$.
Since the local radius of curvature must be $2f$, we find that the point on the OAP surface farthest from the reference plane is at a distance $d' \approx 2f - \sqrt{(2f)^2 - (d/2)^2}$.
Taking $f = 500\,$mm and $d=40\,$mm, this distance turns out to be $d' \approx 0.2\,$mm, which is miniscule compared to the Fresnel length $ d_\mathrm{F} = d^2 / \lambda = 1.6\,$km for $\lambda = 1 \, \mu$m.
Since $d'$ is vastly less than $d_\mathrm{F}$ and the beam is not strongly converging or diverging, the field will not incur appreciable additional diffraction due to the very short propagation between the reference plane and the mirror surface. 
The lack of additional diffraction in this step puts it into the geometrical regime, which allows use of the \emph{pointwise Fourier Transform} (PFT) inside the ASM.\cite{Wyrowski_CurvedSurfaces_JOSAA19, Wyrowski_CurvedSurfaces_SPIE24}
Sec.~\ref{sec: ASM PFT} describes this more fully.

\noindent {\bf reflection at a curved surface:}
VLF employs the Local Plane Wave Interface Approximation (LPIA) to compute the interaction of the vector electromagnetic field with a dielectric multilayer.\cite{Wyrowski_LPIA_AplOpt00}
In the simulations presented in this article, the LPIA models the reflections from silver OAP mirrors with a SiO$_2$ coating (see Table~\ref{table: optical system}).
At each point $p$ on the mirror surface, the LPIA determines the local direction of propagation of the incident field, which is defined as the direction of the phase gradient at $p$.
The electric field vector is then locally decomposed into transverse electric (TE) and transverse magnetic (TM) components, relative to the local plane of incidence at $p$.
The reflected field vector is then computed using the Transfer Matrix Method (TMM), applied to the multilayer stack under the local angle of incidence.\cite{Born&Wolf}
The resulting complex-valued TE and TM reflection coefficients are then used to construct the reflected field vector.
This process is carried out pointwise across the surface according to a spatial sampling scheme.
The magnitudes of the TM and TE coefficients as a function of the incidence angle for the OAP mirrors simulated here are provided in Fig.~\ref{fig: OAP coating}.

\noindent {\bf curved surface-to-plane:} 
This operation is the reverse of the one described just above under ``plane-to-curved surface." 
Once the field on the curved surface has been determined via the LPIA method, it must be propagated back to the reference plane. 
Because there is no standard algorithm for applying the FFT on a curved surface, the conventional ASM with the FFT cannot be used here.  
On the other hand, the PFT can be evaluated on curved surface, making the ASM with the PFT the only viable option for this propagation step, at least within the field-based framework of VLF.
Sec.~\ref{sec: ASM PFT} provides further details.

\subsubsection{The ASM with the PFT for Propagation to a Curved Surface}\label{sec: ASM PFT}
Most applications of the ASM employ the FFT, but for the propagation between the reference plane and the curved surface, VLF employs ASM with the \emph{pointwise Fourier transform} (PFT), henceforth denoted as the \emph{ASMPFT}.
This section outlines the application of the ASMPFT for the purpose of understanding the simulations presented here.  
Those seeking more mathematical rigor should consult the literature.\cite{Baladron_HomeomorphicFT,Wyrowski_CurvedSurfaces_JOSAA19,Wyrowski_Seamless_SPIE24}

The ASMPFT and ray tracing methods (not available in VLF) have several commonalities:
\begin{itemize}
\item As geometrical methods, and they do not account for diffraction.
\item They are \emph{pointwise} and therefore require $\mathcal{O}(N)$ operations in which $N$ is the number of sample points on the originating surface.  In the case of ray tracing, there are $N$ rays that sample the surface.  In the case of the ASMPFT the $N$ sample points are not associated with rays, but are the origination points of $N$ independent calculations.
\item Both methods have a vectorial direction associated with the $N$ sample points.  In the case of ray tracing, this direction must be prespecified, but for the ASMPFT, it is taken to be direction of the gradient of the phase.
\end{itemize}

In order to explain the short free-space propagation between the reference plane and the curved surface, it may help to recall how the classical ASM algorithm employs the FFT to propagate the field between parallel planes:\cite{IntroFourierOptics} The first step is to calculate the angular spectrum of the field, which is a scaled version of the FT, via the FFT algorithm.
The second step is to multiply the angular spectrum by a phase factor that is proportional to the propagation distance $z$, resulting in the angular spectrum at the destination.
The final step is recover the field from its angular spectrum with the inverse FT via the IFFT algorithm.

Consider a complex-valued, monochromatic (wavelength $\lambda$) scalar field $f(x,y)$ and its FT $\tilde{f}(k_x, k_y)$, where $x$ and $y$ are the spatial coordinates, and $k_x$ and $k_y$ are the spatial frequency coordinates.
The angular spectrum (AS) of $f$ is $\tilde{f}(\alpha / \lambda, \beta / \lambda)$, where the direction cosines are $(\alpha ,\beta ) = (\lambda k_x, \lambda k_y )$.  
If $f(x,y)$ is plane wave traveling in the direction $(\alpha_0, \beta_0 , \sqrt{1 - \alpha_0^2 - \beta_0^2})$, its AS has only a single non-zero component: $\tilde{f}(\alpha / \lambda, \beta / \lambda)\delta(\alpha - \alpha_0) \delta(\beta - \beta_0)$, where $\delta(\,)$ is the Dirac delta.

A field $f(x,y)$ in the geometrical regime has a single travel direction at any point $(x,y)$, taken to be the direction of the gradient of its phase, and the corresponding direction cosines are denoted as $\big(\alpha(x,y), \beta(x,y)\big)$.
Since the spatial frequencies and the direction cosines are related via $(\alpha ,\beta ) = (\lambda k_x, \lambda k_y )$, one can determine the direction cosines at the $N$ sample points 
and thereby specify sampled representations of the functions $k_x(x,y)$ and $k_y(x,y)$.
These functions map the $x-y$ plane into the $k_x-k_y$ plane and allow the PFT to be represented as a sum over the $N$ sample points, which is an $\mathcal{O}(N)$ task, as compared to $\mathcal{O}(N) \ln(N)$ for the FFT.

In addition to the $\mathcal{O}(N)$ computational effort required, the ASMPFT readily allows propagation to a curved surface.
Take the source plane to be located at $z=0$ and the destination surface to be specified by the function $z(x,y)$.
Then, the AS associated with the point $(x,y,0)$ is \newline $\tilde{f}\big(\alpha(x,y) / \lambda, \beta(x,y) / \lambda \big)\delta(\alpha - \alpha_0) \delta(\beta - \beta_0)$.
The AS associated with destination surface at the point $\big(x,y,z(x,y)\big)$ is then given by the usual formula:\cite{IntroFourierOptics}
\begin{equation}
\tilde{f}\left(\frac{\alpha}{\lambda}, \frac{\beta}{\lambda}, z\right) =
\tilde{f}\left(\frac{\alpha}{\lambda}, \frac{\beta}{\lambda }, 0\right)
\exp\left[j \frac{2 \pi }{\lambda} z \sqrt{1 - \alpha^2 - \beta^2 \,}   \right] \, ,
\label{eq: AS PFT}
\end{equation}
where $\alpha = \alpha(x,y)$, $\beta=\beta(x,y)$ and $z=z(x,y)$.
Once the angular spectrum has been determined on the surface in this way for each of the $N$ sample points, the field on the surface, $f\big(x, y, z(x,y) \big)$ is found by applying the \emph{inverse pointwise FT} (IPFT), which is based on the same principles as the PFT.\cite{Wyrowski_CurvedSurfaces_JOSAA19, Wyrowski_CurvedSurfaces_SPIE24}

\section{Results}\label{sec: results}

With all of the methods used to model the fields and intensities in the coronagraph's detector plane described in the previous sections, this section is devoted to providing examples of consequences of the coherent nature of the primary and secondary intensities.
The coronagraph in this article exhibits symmetries leading a single primary intensity and a single secondary intensity as per Eq.~\eqref{eq: detector intensity symmetric}.  The examples below are illustrative of the tightly coupled behavior of the primary and secondary intensities, in contrast to the near invariance of a planetary signal.

\begin{figure}[ht]
	\hspace{-8mm}
	\begin{tabular}{r l}
		\includegraphics[width=.55\textwidth,height=.55\textwidth]{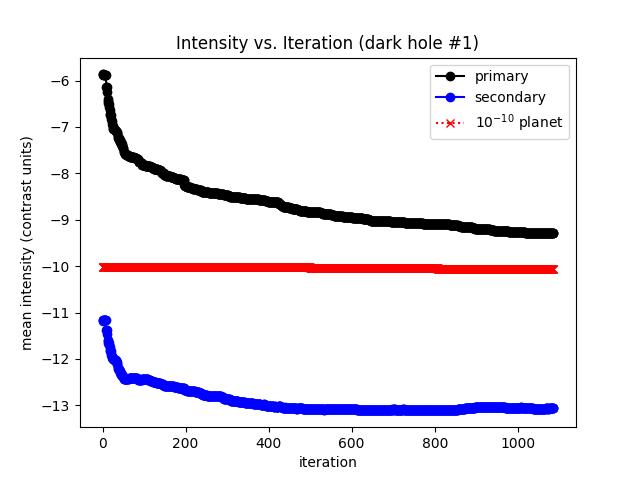} & \hspace{-10mm}
		\includegraphics[width=.55\textwidth,height=.55\textwidth]{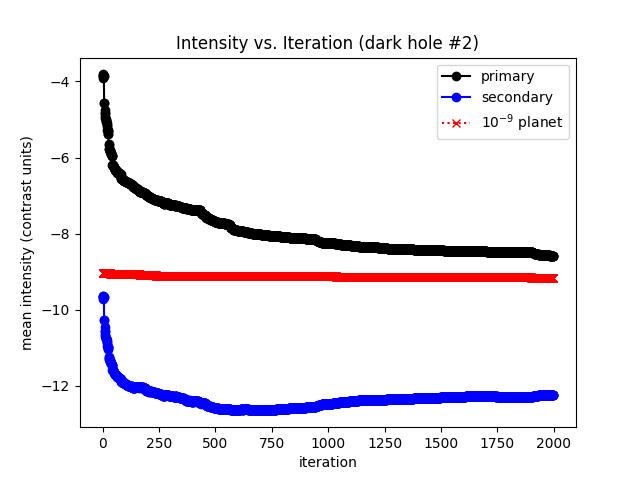}
	\end{tabular}
	\caption{\small Intensities (contrast units) in dark holes \#1 and \#2 as a function of the constrained conjugate gradient (CG) iteration (see Sec.~\ref{sec: sim DH} and Fig.~\ref{fig: All Holes}).  The black curves represent mean primary intensity in the dark hole, and the blue curves represent the mean secondary intensity in the same region.  The red curve is the maximum intensity of an incoherent off-axis source, which plays the role of a planet, placed in the center of the dark hole (see Fig.~\ref{fig: planet PSF}).
	The CG iterations improve the primary dark hole by roughly 4 orders-of-magnitude monotonically, and the secondary dark hole by about 2 orders-of-magnitude non-monotonically.  The off-axis intensity is nearly indifferent to the DM commands corresponding to these iterations, as would be expected of an incoherent off-axis source. 
		\emph{Left:} Dark hole \#1.
		\emph{Right:} Dark hole \#2.
	}
	\label{fig: DH iterations}
\end{figure}

\clearpage

\subsection{Example 1: Simultaneous Dark Holes}\label{sec: sim DH}

\begin{wrapfigure}[18]{r}{.5\textwidth}
	\vspace{-3mm}
	\includegraphics[width=0.5\textwidth]{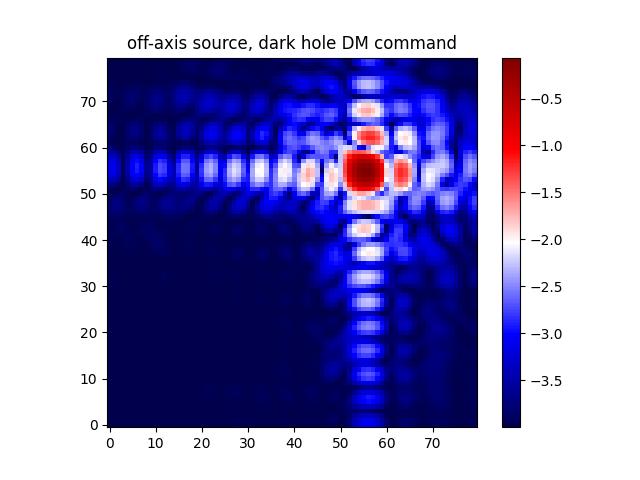}
	\caption{\small Intensity (PSF) of an off-axis source located at the center of dark hole \#1 (see Fig.~\ref{fig: All Holes}).  The non-central portion is somewhat distorted by the DM command corresponding to dark hole \#1.}
	\vspace{-2mm}
	\label{fig: planet PSF}
\end{wrapfigure}

One important result from this study is that a DM command that makes a dark hole in the primary intensity, which is objective of EFC procedures, is likely to also significantly decrease the secondary intensity in the dark hole region.  
This feature may well lessen the design constraint on polarization in future coronagraphs.
It is important to emphasize that the dark holes simulated in this article are achieved via gradient-based minimization of the primary intensity in the desired region.
Any reduction in the secondary intensity is entirely incidental, essentially gratis.

\begin{figure}[t!]
	\includegraphics[width=.99\textwidth]{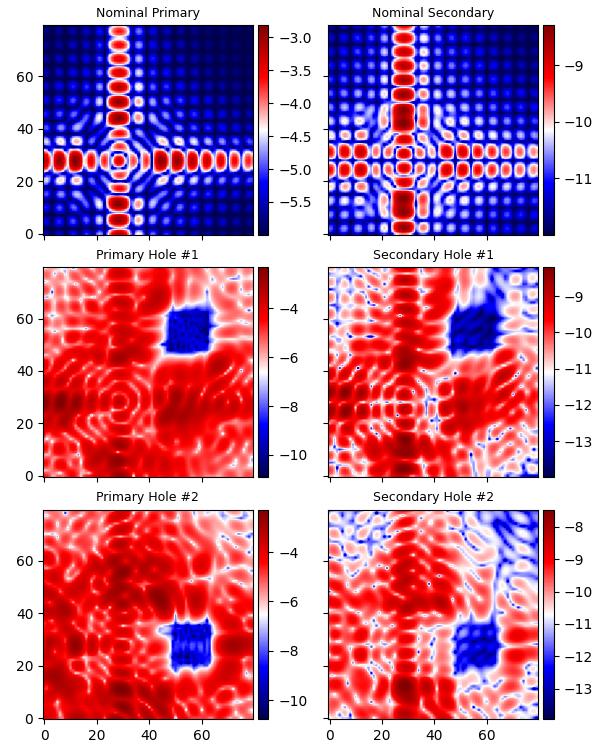}
	\caption{\small PSFs in the Primary and Secondary Intensities with and without dark holes.  The color scales correspond to the intensity in contrast units.  \emph{Left column:} Primary intensity.  \emph{Right column:} Secondary intensity.  \emph{Top row:}  Nominal PSF (same as Fig.~\ref{fig: PSFs}).  \emph{Middle Row:}  PSF with dark hole \#1.  \emph{Bottom Row:} PSF with dark hole \#2.} 
	\label{fig: All Holes}
\end{figure}

The two dark holes simulated here, displayed in Fig.~\ref{fig: All Holes} are both square with an area of $3\times3 (\lambda/D)^2$, located at distance of $4 \lambda/D$ from the center of the detector plane.  Since this coronagraph has no unknown wavefront error, there is no need for alternating sensing and control steps as is usual for EFC implementations.  Indeed, with no unknowns finding dark hole solutions is only a matter of applying standard optimization methods to minimize the intensity in a given region of the detector plane.  Precisely, each dark hole was found by minimizing the mean primary intensity in the specified region, under the constraint that the phase $|\mathrm{difference}|$ between any two actuators is less than $\pi/3$.
The latter was imposed to ensure accurate representation of the DM phasor (see Sec.~\ref{sec: DM}).  A trust-region method that largely relies on conjugate gradient (CG) steps, implemented in {\tt scipy.optimize}, carried out the minimizations.

Fig.~\ref{fig: DH iterations} shows the progress of the primary, secondary and planetary intensity as a function of the iteration number in the minimization process.
It is critical to stress that the minimizer is tasked only with minimizing the primary intensity in the dark hole region, which corresponds to the black curves in the figure.
The black curves decrease monotonically, which is to be expected of gradient-based minimization methods. 
In both cases the minimizer succeeds in decreasing the mean primary intensity in the dark hole by roughly 4 orders-of-magnitude.
Due to the secondary field's coherence with the primary field, the sequence of DM commands that finally reduce the primary intensity by roughly 3 or 4 orders-of-magnitude, incidentally reduce the mean secondary intensity by about 2 orders-of-magnitude, but in a non-montonic manner.
The non-monotonicity of the secondary intensity is hardly perplexing because the minimizer does not take it into account.
In contrast to the primary and secondary intensities, the maximum intensity of the incoherent off-axis source centered in the dark hole, represented by the red curve, is rather oblivious to the changes in the DM corresponding to the dark hole iterations.
This is to be expected because these DM commands do not terribly degrade the Strehl ratio.\cite{Gladysz10}

Fig.~\ref{fig: All Holes} is a closeup of the detector plane, including the center but set towards the upper-right (the $+x, +y$ direction).  
The left column displays the primary intensity and the right displays the secondary intensity, both in constrast units.
In the top row, one can see the nominal (i.e., zero DM command) PSFs, which are no different (other than the zoom) than already shown in Fig.~\ref{fig: PSFs}.
The middle row shows dark hole \#1, located at angle of $45^\circ$ in the plane, whereas the bottom row shows hole \#2, which is located on the x-axis.

\clearpage 

\subsection{Example 2: Joint Modulation}\label{sec: sim mod}

\begin{wrapfigure}[16]{l}{.47\textwidth}
	\hspace{-5mm}
	\includegraphics[width=0.55\textwidth]{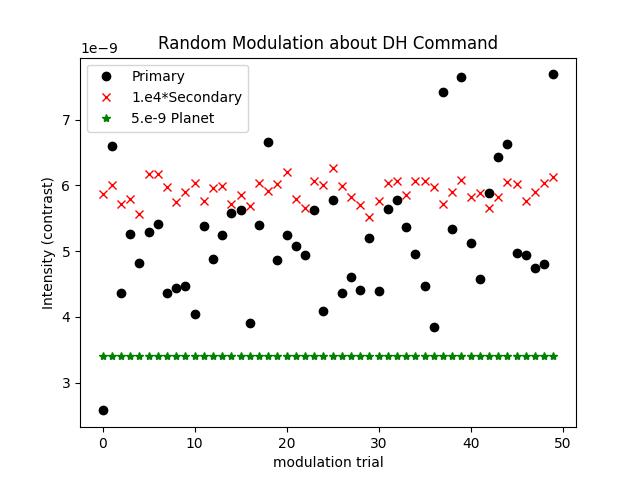}
	\caption{\small The effect of small random DM perturbations on the intensity to simulate the effect of modulation.  The $x$ axis corresponds to the trial number, each of which is a random DM perturbation. 
		Trial \#0, corresponding to leftmost points on the plot, corresponds to the dark hole command itself without perturbation.}
	\label{fig: random modulation}
	\vspace{-0mm}
\end{wrapfigure}

Turning our attention to Fig.~\ref{fig: random modulation}, another consequence of the coherence of the primary and secondary fields is the fact that DM modulations of the primary intensity will also modulate the secondary intensity.
The strategy of using modulation of the DM to separate the starlight from the planet's light is a form of \emph{coherent differential imaging}.\cite{Bottom_CDI_MNRAS17}
The coherence of the secondary field is a possible complication because it has different response to the DM perturbations than does the primary field, and it may need to taken into account.   
However, the fact that a DM command that makes a dark hole in primary intensity may well significantly reduce the secondary intensity, as happens in the examples in this article, makes it more likely that the secondary intensity--- and its modulation--- is negligible.  
Indeed, in the example in Fig.~\ref{fig: random modulation} the secondary intensity values multiplied by $10^4$ in order to appear on the graph. 

A subtlety is fact that a dark hole in primary intensity is the result of a precise construction of destructive interference throughout the dark hole region and the corresponding DM command, let us denote it as $\bc_0$, is close to a minimum of some cost function, even if that function is implicit.  
On the other hand, $\bc_0$ is unlikely to be nearly as close to a minimum of an analogous function of the secondary intensity.
Therefore, one should expect perturbations about $\bc_0$ to be more deleterious to the primary intensity than the secondary one.
This circumstance is observed in Fig.~\ref{fig: random modulation}, in which the dark hole command $\bc_0$ corresponds to trial \#0 (the leftmost data point).   
This figure shows relative modulation of the primary intensity that is larger than that of the secondary intensity.

\section{Summary and Conclusions} 
This article challenges the common parlance that the fields arising from instrumentally induced polarization in a stellar coronagraph are ``incoherent," and it elucidates phenomena that cannot be understood if this descriptor is taken literally.  
The fields and intensities in the detector plane are denoted as either ``primary" or ``secondary," with the former corresponding to what would be obtained in absence of polarization effects (e.g., as modeled with standard Fourier optics methods) and the latter corresponding to the polarization effects of the instrument.
The subject of instrumentally induced polarization is approached by a rigorous analysis consisting of several components:
\clearpage
\begin{enumerate}
	\item The term ``coherence" is defined within the framework of stochastic processes.  
	\item Using the Jones calculus, it is demonstrated that the primary and secondary polarization are fully coherent.  The $90^\circ$ angle between the electric field vectors of the primary and secondary field precludes mutual interference.
	
	\item The linear nature of the coronagraph optical system is leveraged to treat propagation from the entrance pupil, where the DM defines the phase of the input field, to the detector by a single matrix-vector multiplication.  This setup results in 4 matrices, one for each Jones component.
	\item The state-of-the-art VirtualLab Fusion (VLF) package performs all propagation steps in the simulated Lyot-type coronagraph.  VLF provides rigorous and self-consistent treatment of all relevant geometrical, diffraction and vectorial phenomena, including interaction with the materials on the curved OAP surfaces.
	\item Dark holes and random modulations thereof are simulated.
\end{enumerate}

The simulations herein show two examples of dark holes in which the primary intensity is minimized (by $\sim 10^4$), with a concomitant reduction of the secondary intensity (by $\sim 10^2$), which, while incidental to the optimization, is nonetheless achieved.
While this phenomenon is a consequence of coherence phenomena that are present in any astronomical observation with a stellar coronagraph, the properties of the instrumental polarization and the amount of mitigation one should expect need to be modeled on a case-by-case basis. 

This serendipitous reduction of the unwanted secondary intensity is potentially far reaching,
perhaps permitting reduced design constraints in the polarization properties of the optical system or making the requirements for polarization calibration less stringent than otherwise would be the case.
Given the difficulty of the objectives for a next generation space coronagraph, including polarimetry of exoplanets at high contrast in order to uncover biosignatures,\cite{LifePolarimetry_2025} this development should be welcome.

The second consequence of the coherence of the secondary fields is that they are modulated concomitantly with the primary fields (see Fig.~\ref{fig: random modulation}). This arises naturally from their shared dependence on the same DM command vector.
It is likely that modulating the primary intensity through a sequence of DM perturbations, called coherent differential imaging, will be necessary to distinguish the (incoherent) planetary signal from the (coherent) starlight. This strategy relies on the fact that the core of the planetary image is relatively insensitive to DM-induced perturbations, whereas the starlight at the planet's location is not.
Treating the secondary intensity as incoherent would assume an insensitivity to modulation similar to that of the planet, but this is incorrect. Since the secondary field is coherent, it responds to the DM perturbations as well.
In regimes where the secondary intensity is non-negligible, it cannot be assumed to be constant-- its modulation must be taken into account.

Quantitative evaluation of these effects in real systems--namely, the resulting suppression and modulation of secondary intensity in the dark hole--will require modeling efforts similar to those presented in this article but adapted to the specific hardware of each system.
Such efforts would likely involve input beams already carrying polarization and wavefront aberration upon entering the coronagraph and surface errors internal to the coronagraph itself.  
Because these simulations may require thousands of end-to-end propagations, spanning both DM command iterations and realizations of random surface error, the adequacy and computational efficiency of the propagation algorithms will be of paramount importance.
With its efficient, state-of-the-art propagation methods, including the ability to treat curved surfaces with $\mathcal{O}(N)$ operations, VirtualLab Fusion(VLF) stands out as a leading candidate to facilitate such studies.

\subsection*{Data Availability Statement}
The Python codes and supporting data to reproduce the results given in this article are publicly available on GitHub at \emph{https://github.com/ColdStrayPlanet/Optics/tree/master/EFC}

\subsection*{Acknowledgments}
The author would like to acknowledge Frank Wyrowski, Davis Marx, John Kohl, and N. Jeremy Kasdin for helpful discussions.
This work was funded by the Heising-Simons Foundation (grant numbers: 2020-1826, 2022-3912) and 
the National Science Foundation (award number: 2308352).

\bibliographystyle{spiejour} 
\bibliography{../../JOSAA/framework/exop.bib}
\end{document}